\documentstyle[aps]{revtex}

\input{epsfig}

\begin{document}

\textheight 8.8in

\textwidth 6.5in

\topmargin -.25in

\oddsidemargin -.25in

\evensidemargin 0in

\baselineskip 14pt

\def\hm{\ \rm {\it h}^{-1} Mpc}

\def\ltsim{\, \lower2truept\hbox{${<
      \atop\hbox{\raise4truept\hbox{$\sim$}}}$}\,}

\def\gtsim{\, \lower2truept\hbox{${>
      \atop\hbox{\raise4truept\hbox{$\sim$}}}$}\,}

\title{Constraints on flat cosmologies with tracking Quintessence from
  Cosmic Microwave Background observations}

\author{Carlo Baccigalupi$^{1}$\footnote{bacci@sissa.it}, Amedeo
  Balbi$^{2}$\footnote{balbi@roma2.infn.it}, Sabino
  Matarrese$^{3}$\footnote{matarrese@pd.infn.it}, Francesca
  Perrotta$^{1,4}$\footnote{perrotta@sissa.it}, Nicola
  Vittorio$^{2}$\footnote{vittorio@roma2.infn.it}}
\address{$^{1}$ SISSA/ISAS, Via Beirut 4, 34014 Trieste, Italy\\
  $^{2}$ Dipartimento di Fisica, Universit\`a di Roma `Tor Vergata',
  and INFN, Sezione di Roma II\\
  via della Ricerca Scientifica 1, 00133 Roma, Italy \\
  $^{3}$ Dipartimento di Fisica `Galileo Galilei', Universit\`a di
  Padova, and INFN, Sezione di Padova\\
  via Marzolo 8, 35131 Padova, Italy\\
  $^{4}$ Osservatorio Astronomico di Padova, Vicolo dell'Osservatorio
  5, 35122 Padova, Italy}

\baselineskip 10pt \maketitle
\begin{abstract}
  We constrain cosmological parameters in flat cosmologies with
  tracking dark energy (or Quintessence) using the existing data on
  Cosmic Microwave Background (CMB) anisotropies.  We perform a
  maximum likelihood analysis using combined data from COBE/DMR,
  BOOMERanG, DASI and MAXIMA, obtaining estimates for the dark energy
  density $\Omega_{Q}$ and equation of state $w_{Q}$, the physical
  baryon density $\Omega_{b}h^{2}$, the scalar perturbation spectral
  index $n_{S}$, the ratio $R$ between the tensor and scalar
  perturbation amplitude (or the tensor spectral index $n_{T}$).
  
  Dark energy is found to be the dominant cosmological component
  $\Omega_{Q}=0.71^{+0.05}_{-0.04}$, with equation of state
  $w_{Q}=-0.82^{+0.14}_{-0.11}$ ($68\%$ C.L.). Our best fit value of
  the physical baryon density is in good agreement with the primordial
  nucleosynthesis bound.  We find no significant evidence for
  deviations from scale-invariance, although a scalar spectral index
  slightly smaller than unity is marginally preferred. Finally, we
  find that the contribution of cosmological gravitational waves is
  negligible.
  
  These results confirm that Quintessence is slightly preferred with
  respect to ordinary cosmological constant by the present CMB data.
\end{abstract}

\section{Introduction}
\label{intro}
Accurate measurements of the Cosmic Microwave Background (CMB)
anisotropy on sub-degree angular scales represent one of the greatest
achievements in modern cosmology. Two balloon-borne experiments
BOOMERanG \cite{BOOM}, MAXIMA \cite{MAX}, and the ground-based
interferometer DASI \cite{DASI} provided data on the CMB anisotropy
angular power spectrum at multipoles corresponding to angular scales
extending far below the degree, up to a few arcminutes. 
These results give statistically strong evidence for one peak in the 
power spectrum on angular scales corresponding to a degree in the sky 
and significant indications for a second and a third peak on 
smaller angles, confirming and extending earlier 
and less accurate data from BOOMERanG and MAXIMA \cite{CMBEARLIER}, 
Together with the plateau 
observed by COBE/DMR \cite{COBE} on larger angular scales, these CMB
data favour a flat Friedmann Robertson Walker (FRW) cosmological model
and strongly support the existence of super-horizon, almost
scale-invariant curvature perturbations at decoupling, which oscillate
coherently after horizon crossing; this is consistent with a
primordial phase of accelerated expansion as predicted in the context
of the simplest inflationary cosmology \cite{INFLATION}.

The CMB anisotropy is strongly sensitive to the amount of baryons in
the universe, especially through the relative amplitude of acoustic
peaks.  Measurements of the physical baryon density $\Omega_{b}h^{2}$
from the CMB (where $\Omega_{b}$ is the ratio of baryons to critical
density today, and $h$ is the present Hubble parameter $H_{0}$ in
units of 100 km/sec/Mpc) are consistent 
with the Big Bang nucleosynthesis (BBN) \cite{NUCLEO}. 
This indicates that baryons 
can only account for roughly 5\% of the critical density. In fact,
several evidences, including CMB anisotropy measurements, suggest that
the bulk of the total energy density is in some form of ``dark''
non-baryonic component, with Cold Dark Matter particles (CDM)
contributing roughly 25\% of 
the critical density, while about 70\% of the total energy density is
made of a smooth component with negative equation of state. The 
latter, which is the subject of the present work, and is described in
detail below, has attracted a lot of interest in recent years and is
commonly known as ``dark energy'' or ``Quintessence''.

There are at least three independent evidences in favour of dark
energy.  First, Type Ia Supernovae observations indicate that the
universe is experiencing a phase of accelerated expansion
\cite{PERL,RIESS}; recently it has been also noticed that acceleration
is a relatively recent occurrence in the cosmological evolution
\cite{TURNER}.  In FRW cosmologies, cosmic acceleration is possible
only if a component with equation of state less than -1/3 is
dominating the expansion.  Second, best fits of the present CMB data
favour a total energy density which is very close to the critical
value \cite{BOOM,MAX,DASI}.  Third, large scale structure observations
suggest a universe with a low density of clustered material \cite{LSS}.

On the theoretical side, justifying the observed amount of vacuum
energy is extremely difficult. Without entering in a detailed
discussion of the cosmological constant problem, which is probably the
greatest mystery in modern fundamental physics \cite{CCP}, we mention
here the main aspects of this difficulty. If one tries to interpret it
as a vacuum expectation value of some fundamental quantum field, any
known scale of particle physics is tens of order of magnitude larger
than the observed one, up to 120 orders of magnitude in the case of
the Planck scale, leading to an evident ``fine-tuning'' problem.
Moreover, this extremely low level of vacuum energy is such that it is
dominating the cosmic expansion right now, leading to a
``coincidence'' or ``why now'' problem. 
The interest toward dark energy or Quintessence models, first
introduced in \cite{RP,W}, resides in their potential ability
to alleviate these fine-tuning problems, at least at classical level.
Quintessence is the simplest generalization of the
cosmological constant, involving a scalar field $\phi$ with potential
$V$, which provides the required amount of vacuum energy today.
Scenarios with inverse power-law potentials
\cite{RP,INVEXPO}, interestingly connected with high energy particle
physics models \cite{PINV}, have been proven to admit the
existence of ``tracking" solutions in which the dark energy is able to
reach the required value today starting from a very wide set of
initial conditions in the remote past, thus removing the previously
mentioned fine-tuning problem \cite{RP,TRACK},
at least for what concerns cosmological classical trajectories.
Scenarios with exponential potentials \cite{W,INVEXPO},
suggested by string theories \cite{PEXPO},
have been demonstrated to possess ``scaling" solutions in which
the scalar field energy density scales as the dominant cosmological
component, either matter or radiation. Recently, Extended 
Quintessence models \cite{EQ}, in which the dark energy possesses an 
explicit coupling with the Ricci scalar, have been studied
\cite{UZANNMC,AMENMC,CHIBANMC,BP}; a detailed study of tracking
trajectories and of their effects on CMB and Large Scale Structure can
be found in Ref.\cite{TEQ}. In the next Section we briefly recall how
a Quintessence component induces its main effects on CMB spectra.

It is therefore interesting to study how the measured CMB anisotropy
constrains Quintessence models. This has been done in several ways in
the past, by considering earlier data from MAXIMA and BOOMERanG
\cite{CMBEARLIER}. CMB anisotropies in (Extended)
Quintessence models have been extensively studied in \cite{EQ,TEQ},
where the CMBFAST \cite{SZ} code for the computation of cosmological
perturbations was upgraded to include scalar-tensor theories of
Gravity, both for the background and the perturbations, 
in full generality. In a
previous work by us \cite{BBMPV}, a minimally coupled Quintessence
with inverse power law potentials in the tracking regime, was assumed,
to obtain constraints on the Quintessence energy density and its
equation of state: we found $0.3\ltsim\Omega_{Q}\ltsim 0.7$ and
$-1\ltsim w_{Q}\ltsim -0.6$ at $95\%$ confidence level, favouring
potentials $V(\phi )\propto \phi^{-\alpha}$ with $\alpha\ltsim 2$. In
Ref.\cite{AME}, limits on the coupling between Quintessence and dark
matter have been obtained. In \cite{BRAX} constraints from exponential
and inverse power law potentials have been compared.  More recently
\cite{WETTE,WETTECMB,CORASANITI} the impact on CMB anisotropies of the
spectral separation of acoustic peaks has been studied.

In this work we consider the most recent data from BOOMERanG, MAXIMA,
and DASI, and we relax many of the assumptions made in \cite{BBMPV} on
the underlying cosmological model, deriving constraints not only on
the Quintessence parameters but also on the abundance of cosmological
gravitational waves, perturbation spectral indices, and physical
baryon density $\Omega_{b}h^{2}$.  We assume Gaussian and
adiabatic initial conditions (see, however, Ref.\cite{PB}) and a flat
geometry.

Before ending this Section we wish to emphasize that our results on
Quintessence parameters have to be interpreted in a general way. In
fact, as discussed in the next Section, tracking solutions with
inverse power law potentials predict a nearly constant equation of
state at redshift where the Quintessence is important: the latter case
coincides with the simplest model of Quintessence, where this is
assumed to consist of a smooth component with constant $w_{Q}\ge -1$,
independently of its nature.  Therefore, before going to the specific
potential parameters, our results on $w_{Q}$ constrain the time
variation of the vacuum energy from a general point of view.

The paper is organised as follows. In Sect.II we briefly review the
theoretical properties of cosmologies with dark energy. In Sect.III we
describe the region in parameter space which we investigate. In
Sect.IV we show the results of the likelihood analysis of CMB data.
Sect.V contains a discussion of our results.

\section{Tracking Quintessence and CMB anisotropies}

In this Section we review the relevant aspects of tracking
Quintessence scenarios, 
highlighting the scalar field dynamics 
and the main effects on CMB anisotropies. We use our numerical 
code which is a modified version of CMBFAST \cite{SZ}, 
integrating scalar field cosmological 
equations for background dynamics as well as linear perturbations, 
in the general context of scalar-tensor theories of gravity, 
see \cite{EQ,TEQ} for details. 

\begin{figure} 
\begin{center}
  \epsfig{file=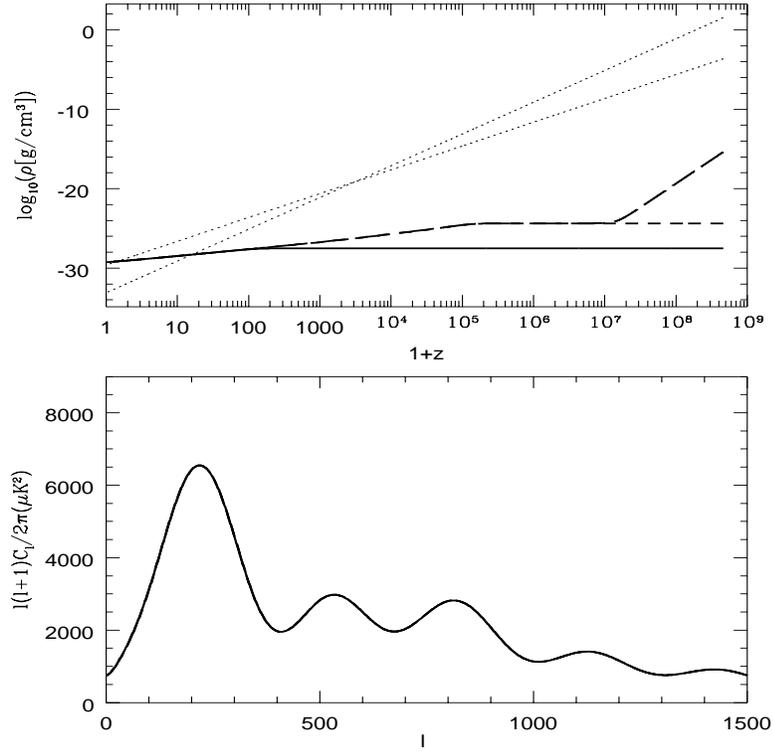,height=4.in,width=5.in}
\end{center}
\caption{Top panel: different Quintessence trajectories converging 
  to the same tracking regime (solid, long and short dashed curves);
  matter and radiation are represented by the dotted curves.  Bottom
  panel: CMB spectra, nearly identical, for the three trajectories.}
\label{f1}
\end{figure}

We restrict our analysis to minimally coupled Quintessence. By 
considering the conformal time $\tau$ in a FRW metric, the 
Fridemann and Klein Gordon equations
for the evolution of the scale factor $a$ and for the dark energy
field $\phi$ take the form
\begin{equation} 
\label{FriedmannKG}
{\cal H}^{2}=\left({\dot{a}\over a}\right)^{2}= {8\pi G \over 3} \left(
a^2 {\rho}_{fluid} +{1 \over 2} \dot{\phi}^2 + a^2 V \right) 
\ ,\ \ddot{\phi}+2{\cal H} \dot{\phi}= -  a^2 V_{\phi}  \ ,
\end{equation}
where the dot and the subscript $\phi$ denote differentiation with
respect to the conformal time $\tau$ and to the Quintessence scalar
field, respectively, while $fluid$ represents contributions from all
the species but Quintessence and $V(\phi)$ is the scalar field
potential.  We will generally describe the amount of a particular
species $x$ with the present-day ratio $\Omega_{x}$ of its energy
density $\rho_{x}$ to the critical one 
$\rho_{c}=3({\cal H}/a)^{2}/8\pi G$.  
As we already mentioned, several potential
shapes are under study: cosine \cite{COS}, exponential \cite{PINV} and
inverse power law \cite{PEXPO}.  The cosine type has been recently
considered in the context of Extended Quintessence \cite{CHIBA}. With
an exponential type potential, dark energy possesses the same redshift
dependence as the dominant component which is driving the cosmic
expansion, either matter or radiation, and some ad-hoc mechanism or a
non-tracking behavior is required in order to bring it to domination.
In the inverse power law case, the tracking solution possesses an
approximately constant equation of state which is not set by the
component which leads the cosmic expansion but by the potential
exponent itself. We take therefore 
\begin{equation}   
V(\phi )={M^{4+\alpha}\over\phi^{\alpha}}\ ,
\label{v}
\end{equation}
where the value of $\alpha >0$ will be specified later and the
mass-scale $M$ is fixed by the level of energy contribution today 
from the Quintessence field. 
As we did in our previous works \cite{EQ,TEQ,BBMPV}, 
we integrate numerically equations 
(\ref{FriedmannKG}) with the potential (\ref{v})
to get the evolution of cosmological background 
quantities. As first noted in \cite{RP}, the Quintessence scalar 
field joins tracking solutions, which are most simply expressed 
in terms of the Quintessence energy density and pressure:
\begin{equation}
\label{rhoqpq}
\rho_\phi=\rho^{kin}_{\phi}+\rho^{pot}_{\phi}=
{1\over 2a^{2}}\dot{\phi}^{2}+V(\phi )
\ \ ,\ \ p_\phi=\rho^{kin}_{\phi}-\rho^{pot}_{\phi}\ .
\end{equation}
We are interested in Quintessence equations of state in the range
$-1\le w_{Q}\le -0.5$, because this is the typical interval leading to
cosmic acceleration today \cite{PERL,RIESS}; 
since during the tracking regime the equation of state is 
roughly constant in time, 
it follows that the
Quintessence energy density plays a role in the cosmic evolution only
at low redshifts, $1+z\ltsim 10$, since the pressureless matter
density increases much more rapidly with redshift. 
To give an intuitive description of the principal cosmological 
effects of Quintessence, it is enough to describe tracking
dynamics in the Matter Dominated Era (MDE). 
Since during the MDE the
scale factor goes as $a(\tau)\sim\tau^{2}$, it easy to see that a
power law solution $\phi\propto\tau^{\gamma}$ to the Klein Gordon
equation (\ref{FriedmannKG}) exists for
\begin{equation}
\label{KGtracksol}
\alpha ={6\over\gamma}-2\ . 
\end{equation}
In this regime, the Quintessence energy density scales with 
redshift as
\begin{equation}
\label{rhotrack}
\rho_{\phi}\propto (1+z)^{3(1+w_{Q})}
\end{equation}
where its equation of state is
\begin{equation}
\label{wqtrack}
w_{Q}=-{2\over \alpha +2}\ .
\end{equation}
Our numerical integrations reproduce with good approximation 
the tracking conditions (\ref{rhotrack},\ref{wqtrack}), even 
if the equation of state is not perfectly constant and this 
have interesting consequences \cite{WDOT}.
The interesting feature of these solutions, which makes their
importance as tracking trajectories, is that they can be reached
starting from a very wide set of initial conditions, so that the the
initial dark energy density deep in the Radiation Dominated Era (RDE)
could have been between the present value and tens of orders of
magnitude higher \cite{TRACK}. We show an example of this
phenomenology in Figure \ref{f1}. In the top panel we plot the
redshift evolution of matter, radiation and Quintessence starting from
different initial conditions for the model specified below in
Eq.(\ref{bestfit}). Matter and radiation (light dotted curves) have
the known scalings; Quintessence trajectories converge to the same
tracking regime starting from different initial conditions. The
potential parameter is $M\simeq 0.6M_{P}$, where $M_{P}=1/\sqrt{G}$ is
the Planck mass, and $\alpha =-0.8$; the initial conditions for the
solid, short dashed and dashed curve respectively are:
\begin{equation}
\phi_{in}=10^{-3}M_{P}\ ,\ \rho^{kin}_{in}=0
\ ,\ \phi_{in}=10^{-7}M_{P}\ ,\ \rho^{kin}_{in}=0
\ ,\ \phi_{in}=10^{-7}M_{P}\ ,\ \rho^{kin}_{in}/\rho^{pot}_{in}=10^{9}\ .
\end{equation}
These examples give an idea of the stability of tracking solutions.
Since the different trajectories differ only at high redshifts when
Quintessence is subdominant the corresponding CMB spectra, shown in
the bottom panel, are to high precision indistinguishable. 

Cosmological linear perturbations, including Quintessence 
fluctuations $\delta\phi$ obeying the perturbed Klein Gordon 
equation 
\begin{equation}
\label{KGp}
\delta (\Box\phi )+V_{\phi\phi}\delta\phi=0\ ,
\end{equation}
are numerically evolved starting 
from Gaussian adiabatic initial conditions: 
we refer to our previous works \cite{PB,EQ,TEQ} for full details and 
formalism. Here we give only an intuitive description 
of the main effects of Quintessence on CMB, which can be understood 
by considering the tracking behavior of the background evolution as
expressed by equations (\ref{rhotrack}) and (\ref{wqtrack}). 
Note that the tracking regime itself is strictly valid only if matter
dominates the expansion. When Quintessence comes to dominance the
universe accelerates its expansion and the dark energy leaves the
tracking regime. This makes the present equation of state different
from the one during the tracking regime; however, this difference is 
only at the level of $10\%$ for all the cases we consider in deriving 
our constraints. 

As $w_{Q}$ moves from the cosmological constant case ($-1$) to
larger values, the conformal distance $\tau_{dec}$ to the last
scattering surface gets reduced:
\begin{equation}
\label{taudec}
\tau_{dec}=H_{0}^{-1}\int_{0}^{z_{dec}}
{dz\over\sqrt{\Omega_{matter}(1+z)^{3}+\Omega_{Q}(1+z)^{3(1+w_{Q})}}}\ .
\end{equation}
As a consequence, the location of all the acoustic features in the CMB
spectra shifts toward lower multipoles, corresponding to large angular
scales in the sky.

The same mechanism leads to a reduction of the Integrated Sachs-Wolfe
(ISW) effect. The latter is represented by extra power at low
multipoles, $l\ltsim 10$, caused by the dynamics of gravitational
potentials at low redshift, $z\ltsim 10$ \cite{HU}; since the distance
decreases in Quintessence models with respect to cosmological constant
ones, as shown by equation (\ref{taudec}), the gravitational potentials
dynamics is reduced correspondingly.  However, for values of $w_{Q}$
well above $-1$, the Quintessence starts to dominate the cosmic
expansion earlier in time, leading to an enhancement of the time
interval in which the cosmic equation of state changes and therefore
to an increase of the ISW power. As a consequence of these two
competing effects, the ISW effect gets slightly reduced for $-1\ltsim
w_{Q}\ltsim -0.8$, while it increases for larger values.

These two effects are purely geometric, i.e. they do not affect, for a
given level of $\Omega_{Q}$, the shape of the acoustic peaks on
sub-degree angular scales. The latter changes if $\Omega_{Q}$ and the
ratio between CDM and baryons changes.

In the next Section we will discuss in detail the cosmological
parameters that we consider, and the region in the parametric space
which we investigate.

\section{Gridding cosmological parameters}

In this work we improve considerably the extension and the gridding of
the cosmological parameter space with respect to what we did in
Ref.\cite{BBMPV}.

We consider flat cosmologies and fix the value of the Hubble parameter
at present to:
\begin{equation}
\label{h}
H_{0}=65\ {\rm km/sec/Mpc}\ ,
\end{equation}
in agreement with current estimates \cite{HUBBLE}. We then vary the
present ratio of baryon to critical density, in order to obtain
different values of the physical baryon density, $\Omega_{b}h^{2}$. 
The amount of baryons in the universe is one of the
most important quantities which affect CMB acoustic oscillations. At
decoupling, any photon-baryon density fluctuation at wavenumber $k$
which is entering the horizon oscillates between compression and
rarefaction under the effect of the potential wells and hills caused
by the dark matter perturbations. Increasing the baryon amount favours
compression with respect to rarefaction peaks, simply because it
shifts toward the bottom of the potential wells the rest position of
the oscillator \cite{HU}.  As we already stressed, tracking
Quintessence is described by its energy density today and by its
equation of state.  The radiation component is made of photons and
three massless neutrino families; the matter component is made of CDM
plus baryons.

Concerning perturbations, we allow for departure from scale invariance
of the initial perturbation spectrum, as well as for a non-zero
amplitude of cosmological gravitational waves.  Scale invariance is
related to the scalar spectral index, $n_{S}$; the latter can be
defined in terms of the scalar perturbation spectrum at the horizon
crossing as
\begin{equation}
{\cal P}(k)\propto k^{n_{S}-1}\ .
\label{scalar}
\end{equation}
In the following, we will refer to cases with $n_{S}$ larger or
smaller than 1 as ``blue" and ``red" spectra, respectively \cite{MML}.

Cosmological gravitational waves are tensor perturbations of the
metric.  Their power is maximal on super-horizon scales, corresponding
to $\ell\ltsim 200$ on the CMB angular power spectrum, making this
multipole region the only one where this component is detectable. Its
power can be parametrized in terms of the ratio $R$ between the tensor
and the scalar contributions to the CMB quadrupole; moreover, we adopt
the single field inflationary consistent relation (see e.g.
Ref.\cite{INFLATION}), linking the tensor spectral index to $R$.
Summarizing, we define
\begin{equation}
\label{ntr}
R={C_{2}^{tensor}\over C_{2}^{scalar}}\ ,\ n_{T}=-{R\over 6.8}\ ,
\end{equation}
where $n_{T}=0$ means here a scale invariant tensor power spectrum at
horizon crossing. The presence of gravitational waves as well as the
deviation from scale invariance can be related to the standard slow
rolling parameters $\epsilon,\eta$ for single field inflation,
which are defined in terms of the inflaton potential and its first 
two derivatives (see e.g. Ref.\cite{INFLATION}). One has:  
\begin{equation}
\label{epsiloneta}
\epsilon = {R\over 13.6}\ ,
\ \eta = {1\over 2}\left(n_{S}-1+{3R\over 6.8}\right)\ .
\end{equation}
A summary of the values of the cosmological parameters considered in
this work is given in Table \ref{table}.

\section{Measures from experimental data}

In this Section we compare theoretical CMB angular power spectra
corresponding to the models in Table \ref{table} with the most recent
data from BOOMERanG, MAXIMA and DASI, as published in
\cite{BOOM,MAX,DASI}, as well as with the 4 year COBE/DMR data
\cite{COBE}. Our data set consists of 65 data points: 19 from
BOOMERanG in the range $76\le\ell\le 1025$, 13 from MAXIMA in the
range $36\le\ell\le 1235$, 9 from DASI in the range $104\le\ell\le
864$, and 24 from COBE/DMR in the range $2\le\ell\le 25$. As in
\cite{BBMPV}, for a given choice of parameters we compare the measured
quantities $\ell(\ell+1)C_\ell/2\pi$ to their theoretical predictions
by evaluating the likelihood of the data, ${\cal L}\propto
\exp(-\chi^2/2)$.  
When possible (i.e. COBE and DASI), we use the offset-lognormal ansatz
for the shape of the likelihood, as in \cite{BJK}. 
We do not consider correlations among the data points of each
experiment, as they are rather small \cite{BOOM,MAX,DASI}. We take
into account the effect of each experiment calibration uncertainty
(20\%, 8\% and 8\% for BOOMERanG, MAXIMA and DASI, respectively
\cite{BOOM,MAX,DASI}) on the measured power spectrum. 
We do not take into account beam and pointing inaccuracies, since the
details of these uncertainties were not made publicly available by the
teams. However, we believe these effects should not affect much the
released data.

Likelihood curves for each parameter are shown in Figure \ref{f2}.
Each curve is obtained by fixing all the other parameters to their
best fit value. It is evident that the preferred flat cosmological
model involves vacuum energy as the dominant component, at the $70\%$
level. Moreover, our best fit value of the baryon density,
$\Omega_{b}h^{2}=0.022$, is in good agreement with the primordial
nucleosynthesis bounds, $\Omega_{b}h^{2}=0.020\pm 0.002$ (95\% C.L.)
\cite{NUCLEO}. The primordial power spectrum is consistent with scale
invariance, $n_{S}\simeq 1$, although slightly red spectra are
favoured. These results are in good agreement with previous analyses
\cite{BOOM,MAX,DASI}.

A first new result of our analysis is that models with a subdominant
contribution of gravitational waves are favoured. As we already
stressed, the main effect of cosmological gravitational waves is to
enhance the anisotropy power above the degree scale, at multipoles
$l\ltsim 200$. The likelihood curves indicate that such a contribution
is unlikely to be required by the present data.  The relative height
of the CMB acoustic peaks with respect to the plateau region is likely
to be determined mainly by the other relevant parameters, i.e. the
relative abundance of cosmological components, the Hubble parameter,
the scalar spectral index.

However, for the purposes of this work the most important result is
the preference for Quintessence models over cosmological constant
ones, as implied by a clear peak in the likelihood curve at
$w_{Q}\simeq 0.8$. This result confirms and strengthens our previous
findings \cite{BBMPV}: a remarkably similar result, although with
larger confidence regions, was obtained by considering earlier CMB
data release \cite{CMBEARLIER}, as well as a reduced grid in number of
parameters and gridding steps. Here we are obtaining almost the same
measure of the vacuum energy equation of state, but considering the
most recent data as well as allowing for variation of new parameters,
namely the gravitational wave amplitude and the scalar spectral index.
To understand the robustness of this result, it is natural to search
what is the dominant effect of Quintessence compared to the
cosmological constant, directly on the CMB spectrum.  We shall return
to this point below.

\begin{figure} 
\begin{center}

  \epsfig{file=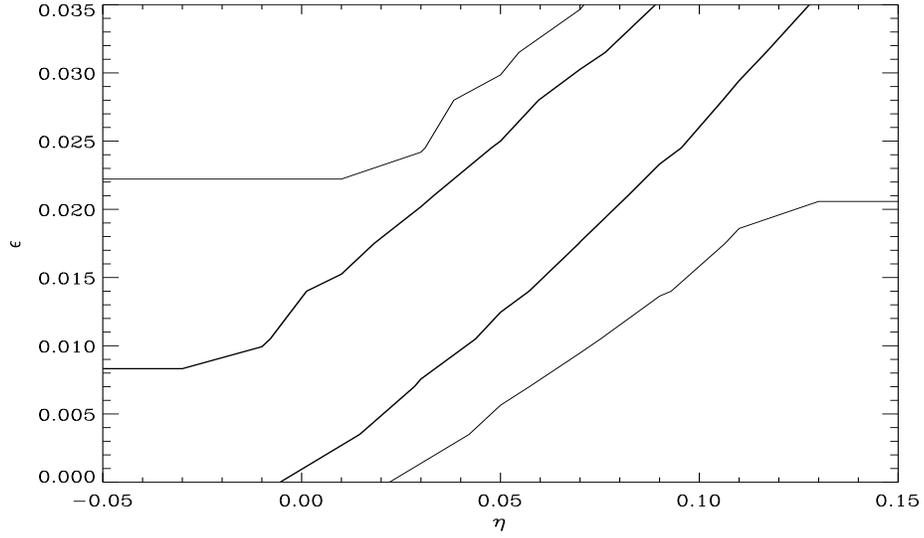,height=3.in,width=5.in,angle=90}
\end{center}
\caption{Likelihood contours at $68\%$ (heavy lines) and $95\%$ 
  (light lines) confidence levels in the plane of the inflationary
  slow rolling parameters $\epsilon,\eta$ defined in equation
  (\ref{epsiloneta}).}
\label{f3bis}
\end{figure}

Figure (\ref{f3}) shows confidence regions for different parameter
combinations.  Heavy and light lines represent $68\%$ and $95\%$
confidence levels, respectively.  The highest degeneracy concerns the
($R,n_{s}$) plane. This degeneracy has a simple explanation.
Increasing the scalar spectral index has the effect of enhancing the
CMB peaks with respect to the region at $\ell\ltsim 200$, and this is
disfavoured by the data. The presence of a tensor component
reintroduces power at low multipoles, so that its net effect, once the
power spectrum is normalized, is to reduce the excess power of
``blue'' models at high $\ell$. The constraints in the $(R,n_{S})$ 
plane, in terms of the parameters $\epsilon$ and $\eta$ defined 
in equations (\ref{epsiloneta}), appear as in Figure \ref{f3bis}. 

\begin{figure} 
\begin{center}
  \epsfig{file=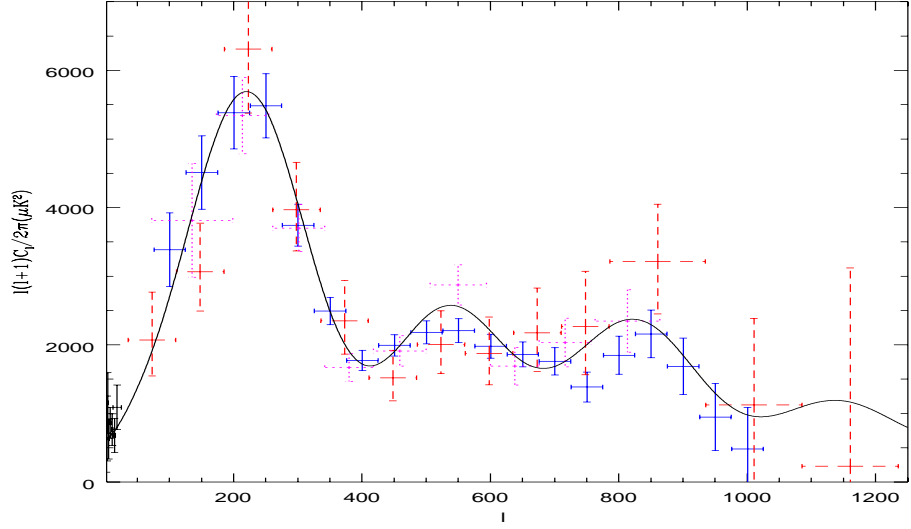,height=3.in,width=5.in}
\end{center}
\caption{Best fit cosmological model vs. experimental data, 
  $\Omega_{Q}=0.72,w_{Q}=-0.8,\Omega_{b}h^{2}=0.022,n_{s}=0.96,R=0$.
  Solid error bars are for COBE (low multipoles) and BOOMERanG data,
  dashed and dotted for MAXIMA and DASI data, respectively.}
\label{f4}
\end{figure}

In all the other cases the contours are closed, at least at $68\%$
confidence level.  Summarizing, our best estimate of the cosmological
parameters considered in this work are, at the 68\% confidence level:
\begin{equation}
\Omega_{Q}=0.71^{+0.05}_{-0.04}\ ,\ w_{Q}=-0.82^{+0.14}_{-0.11}\ ,\ 
\Omega_{b}h^{2}=0.022\pm 0.003
\ ,\ n_{S}=0.95\pm 0.08\ ,\ R\ltsim 0.5\ (n_{T}=-R/6.8)\ .
\label{bestfit}
\end{equation}
We have fit 65 data points using approximately 9 parameters (5
cosmological parameters, 3 calibration parameters and an overall
normalization of the power spectrum) so that the number of degrees of
freedom (DOF) is roughly DOF$\simeq 65-9=56$. Our best fit has
$\chi^2=57$, and $\chi^2/$DOF$\simeq 1$.  The best fit power spectrum,
together with the experimental data, is shown in Figure \ref{f4}.

Before closing this Section, let us translate our limits on the
Quintessence equation of state in terms of the potential slope
$\alpha$ defined in equation (\ref{v}).  As we already stressed, the
result (\ref{bestfit}) concerning the equation of state provides
general evidence in favour of a time variation of the cosmological
vacuum energy; the reason is that tracking trajectories with inverse
power laws correspond to a nearly constant equation of state at the
redshift of interest, which is the simplest model of Quintessence. It
is however interesting to determine the corresponding interval in the
potential slope; tracking trajectories with a present equation of
state as in (\ref{bestfit}) correspond to exponents
\begin{equation}
\alpha = 0.8_{-0.5}^{+0.6}\ .
\label{alphabestfit}
\end{equation}
Note that this is slightly different, although within a $1\sigma$
interval, from what would be obtained from equation
(\ref{wqtrack}), $\alpha =0.7_{-0.5}^{+0.2}$. As we already mentioned,
this is because the result (\ref{alphabestfit}) is numerically
obtained from the present equation of state, which is slightly
different from its value during the tracking regime.

This completes the constraints obtained from the comparison of our CMB
spectra with the data. The most interesting aspect is the evidence in
favour of a time variation of the vacuum energy component.  In the
next Section we will discuss the robustness of these results.

\section{Discussion} 

Our results (\ref{bestfit}) have been obtained under a number of
assumptions.

First, all the perturbations, including those in the Quintessence
component, are assumed to be Gaussian and initially perfectly
adiabatic \cite{PB}.  The tensor spectral index has been related to
the amplitude of gravitational waves through the consistency relation
(\ref{ntr}); moreover, the global geometry is assumed to be flat. Even
if these choices reduce considerably the parameter space region that
we explore, they are justified by the prediction of the simplest
inflationary models \cite{INFLATION}.  In addition, we assumed three
massless neutrino families as well as a dark matter type which is
purely cold; these conditions are those currently preferred by data on
the large scale structure of the Universe \cite{LSS}.

Within these hypotheses, our comparison with CMB data revealed an
interesting indication in favour of a time-varying vacuum energy. Even
if this result depends crucially on the present CMB data which are
still far from the performances that will be reached by satellite
measurements \cite{MAP,PLANCK}, two questions arise.  First, which is
the effect of Quintessence, compared with a pure cosmological
constant, which makes present data prefer a Quintessence component.
Second, which cosmological parameter not considered in this work can
mimic the effects of Quintessence, thus undermining the robustness of
our results.

\begin{figure} 
\begin{center}
  \epsfig{file=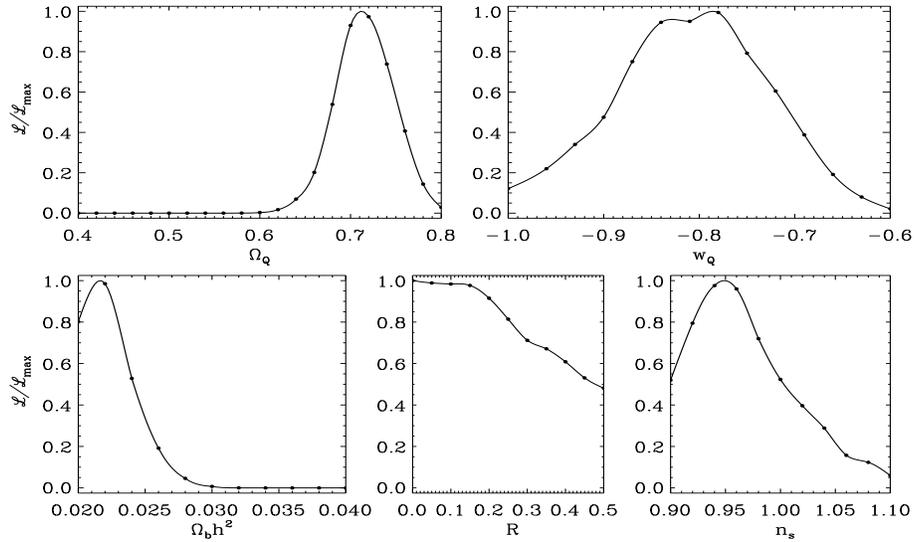,height=3.in,width=5.in,angle=90}
\end{center}
\caption{Likelihood curves for cosmological parameters of Table 
  \ref{table}, normalised to the peak value.}
\label{f2}
\end{figure}

\begin{figure} 
\begin{center}

  \epsfig{file=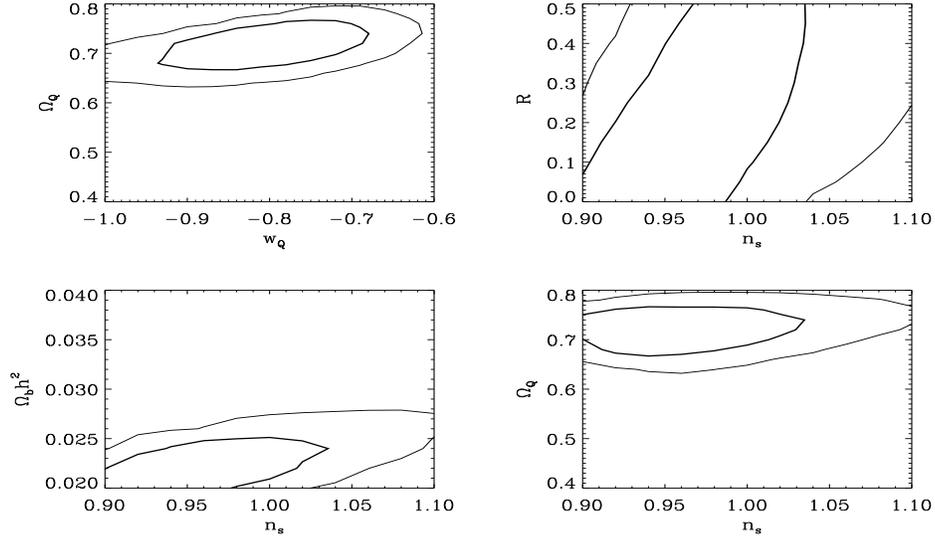,height=3.in,width=5.in,angle=90}
\end{center}
\caption{Likelihood contours at $68\%$ (heavy lines) and $95\%$ 
  (light lines) confidence levels.}
\label{f3}
\end{figure}

\begin{figure} 
\begin{center}
  \epsfig{file=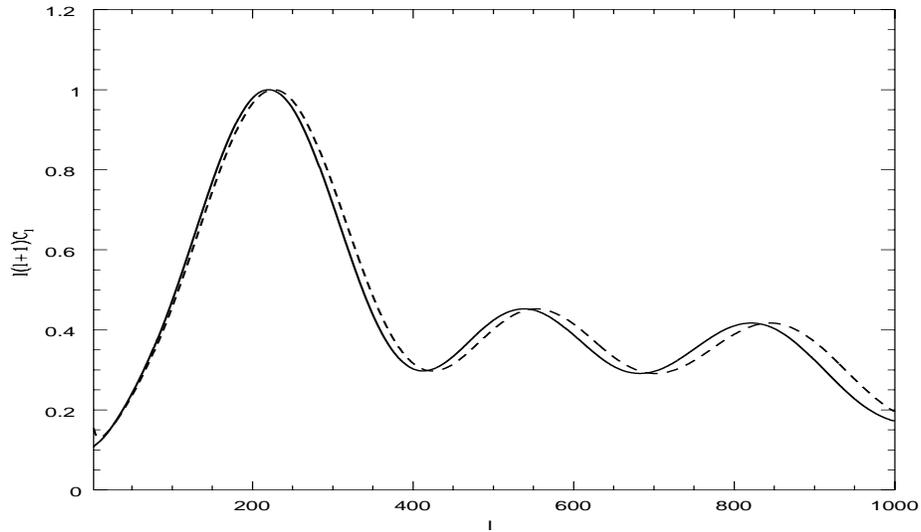,height=3.in,width=5.in}
\end{center}
\caption{First peak normalised CMB spectra for our best fit model 
  ($w_{Q}=-0.8$, solid line) and its cosmological constant equivalent
  ($w_{Q}=-1$, dashed line).}
\label{f5}
\end{figure}

We address the first question by comparing the CMB spectrum which
represents the best fit (\ref{bestfit}) with its cosmological constant
analog, in which all the parameters have the same value except for
the equation of state, which is set to $-1$.  The two spectra are
compared in Figure \ref{f5}, normalised to the first peak height. It
is evident that they differ mainly because Quintessence
causes a systematic shift of all the acoustic features toward larger
angular scales, or lower multipoles; as we already mentioned in
Section II, this is due to a progressive reduction of the conformal
distance between us and the last scattering surface, as $w_{Q}$ moves
from $-1$ to higher values. Note also that the effect is small
compared to the data error bars; indeed the cosmological constant, at
the $95\%$ confidence level, is still compatible with present data, as
it is evident in Figures \ref{f2} and \ref{f3}.

The answer to the second question can now be given. If we fix the
relative abundances today, the only parameter which we do
not consider and that could mimic the projection effect in Figure
\ref{f5} is the cosmic curvature, represented by the total density
parameter $\Omega_{tot}$. It is indeed well known that a closed
universe with $\Omega_{tot}>1$ moves the acoustic features of the CMB
spectrum toward smaller multipoles \cite{HU}. This argument is
supported by earlier constraints on cosmological parameters -- obtained
without considering Quintessence models -- that were set by experiments
measuring sub-degree CMB anisotropies; the quoted results for
$\Omega_{tot}$ were $1.04^{+0.05}_{-0.05}\ ({\rm or}\ 
1.02^{+0.06}_{-0.05})$, $0.90_{-0.16}^{+0.18}$, $1.04\pm 0.06$ for
BOOMERanG, MAXIMA and DASI, respectively \cite{BOOM,MAX,DASI}.
Although flatness is well within $1\sigma$ for all the experiments, it
is however interesting that Quintessence offers a mechanism to explain
the slight preference of the existing data for closed models, the same
mechanism being the reason why we find indications in favour of a
dynamical vacuum energy in this work.

We conclude that, in the framework of flat cosmologies, the preference
in favour of Quintessence from present CMB data is quite robust.
Future satellite missions \cite{MAP,PLANCK} will allow to further test this
result.

\acknowledgements

We used a modified version of CMBFAST \cite{SZ}. We are grateful to 
Fabio Pasian and Claudio Vuerli for support in data storing. 
AB acknowledges fruitful discussions with Domenico Marinucci. 

\begin{table*}
\caption{Cosmological parameters values.}
\begin{tabular}{lcccc}
Parameter & minimum & maximum & step \\
\hline
$\Omega_{Q}$ & $0.40$ & $0.80$ & $0.02$\\
$w_{Q}$ & $-1.00$ & $-0.60$ & $0.03$\\
$\Omega_{b}h^{2}$ & $0.20$ & $0.40$ & $0.02$\\
$\Omega_{CDM}$ & $1-\Omega_{Q}$ & $1-\Omega_{Q}$ & $0.02$\\
$n_{S}$ & $0.90$ & $1.10$ & $0.02$\\
$R$ & $0$ & $0.50$ & $0.05$\\
$n_{T}$ & $-R/6.8$ & $-R/6.8$ & $0.05/6.8$
\end{tabular}
\label{table}
\end{table*}

\end{document}